\documentstyle[12pt,epsfig]{article}
\newcommand{\Frac}[2]%
{{\textstyle \frac{\mbox{\footnotesize $#1$}\rule[-0.9mm]{0mm}{1mm}}%
{\mbox{\footnotesize $#2$}\rule{0mm}{3.1mm}}}}
\renewcommand{\thefootnote}{\fnsymbol{footnote}}

\begin{document}
\begin{titlepage}
\vspace*{-12 mm}
\noindent
\begin{flushright}
\begin{tabular}{l@{}}
TUM/T39-00-10 \\
hep-ph/0008289 \\
\end{tabular}
\end{flushright}
\vskip 12 mm
\begin{center}
{\large \bf 
$g_1$ at low $x$ and low $Q^2$ with Polarized $ep$ Colliders
}
\\[14 mm]
{\bf Steven D. Bass}
\footnote[2]{steven.bass@cern.ch} 
\\[5mm]   
{\em 
Physik Department, 
Technische Universit\"at M\"unchen, \\
D-85747 Garching, Germany}\\[10mm]
{\bf Albert De Roeck}
\footnote[3]{albert.de.roeck@cern.ch}
\\[5mm]   
{\em 
CERN, CH-1211 Geneva 23, Switzerland} \\[10mm]

\end{center}
\vskip 10 mm
\begin{abstract}
\noindent
Measurements of $g_1$ at low $x$ and low $Q^2$ are expected to provide a 
sensitive probe of the transition from Regge to perturbative QCD 
dynamics, offering a new testing ground for models of small $x$ physics.
We discuss the potential of polarized $ep$ colliders (Polarized HERA and
eRHIC) to investigate this physics
--- varying $Q^2$ between 0.01 and 1 GeV$^2$ 
--- and to constrain the high-energy part of 
the Drell-Hearn-Gerasimov sum-rule for polarized photoproduction.
\end{abstract}
\end{titlepage}
\renewcommand{\labelenumi}{(\alph{enumi})}
\renewcommand{\labelenumii}{(\roman{enumii})}
\renewcommand{\thefootnote}{\arabic{footnote}}

\newpage

\baselineskip=6truemm

\section{Introduction}

The HERA measurements of the proton structure function $F_2 (x,Q^2)$ 
at low $x$ and low $Q^2$ (less than 1 GeV$^2$) mark the transition
between Regge ``confinement physics'' and perturbative QCD.
For fixed $Q^2$ up to 0.65GeV$^2$ the measured large 
$s_{\gamma p} \sim Q^2/x$ behaviour of $F_2$ is consistent with 
the expectations of Regge phenomenology \cite{donn}.
For values of $Q^2$ greater than 1GeV$^2$ the $F_2$ data is well 
described \cite{grv} by DGLAP evolution~\cite{dglap} and rises 
faster with decreasing $x$ than the soft Regge prediction.
The ``transition region'' between photoproduction and deep inelastic 
scattering and the onset of perturbative QCD is a subject of much 
current interest \cite{ijmp,bkrmp}.

The small $x$ behaviour of nucleon structure functions $f(x,Q^2)$ 
is often described in terms of an effective intercept $\lambda$
($f(x, Q^2) \sim x^{- \lambda}$ at small Bjorken $x$) 
which changes between 0.1 and 0.4 for unpolarized data. 
In the transition from photoproduction to deep inelastic values of
$Q^2$ much larger changes are expected in the effective intercept 
for the $g_1$ spin dependent structure function compared to 
the spin independent structure function $F_2$ -- see Section 2 below.
In this paper we discuss the potential of Polarized $ep$ Colliders
(Polarized HERA  and eRHIC)
\cite{phera,erhic,epjc} to measure
the spin dependent part of the total photon--nucleon  cross-section 
for photon virtualities $Q^2$ between 0 and 1 GeV$^2$.
These measurements would help constrain the high-energy part of the
Drell-Hearn-Gerasimov sum-rule \cite{dhg} for spin dependent 
photoproduction and impose new constraints on theoretical models 
which aim to describe the transition from soft to hard physics at 
small $x$.
Open questions are: 
At which $Q^2$ does the effective intercept for $g_1$ start to grow ? 
What is the rate of growth with increasing $Q^2$ ? 
Where in $Q^2$ will perturbative QCD start to describe future $g_1$
data at small $x$ ?

The physics program of the
Polarized HERA project is documented in \cite{phera} and 
Polarized eRHIC is discussed in \cite{erhic}. HERA $ep$ collisions 
are at a centre of mass (CMS) energy 300 GeV, while eRHIC is foreseen
to operate at a CMS energy of 100 GeV.
The low $Q^2$ measurements would complement deep inelastic 
measurements of $g_1$ at low $x$ (extend to low $Q^2$) and
studies of $\Delta g$ and spin dependence of diffraction.
The expected electron and proton beam polarizations are 
$P_e = P_p = 70 \%$ and the
integrated luminosity is ${\cal L} = 500$pb$^{-1}$ (HERA) or 
${\cal L} = 4$fb$^{-1}$ (eRHIC) after several years of data taking. 
Hence  one expects 
\cite{epjc} to be able to measure the electron-proton spin asymmetry at small
$x$ (less than about 0.05) to a precision $\delta A \simeq 0.002 - 0.0001$ 
or better in 
deep inelastic scattering. 
Recent ideas on using polarized deuterons may give access  to 
$g_1^p$ and $g_1^n$ separately.

In Section 2 we outline the key physics issues. 
Experimental aspects are discussed in Section 3. 
Section 4 contains an estimate of the possible asymmetries. 
Finally, in Section 5 we conclude.

\section{The transition region}

We first recall what is known about the transition region in $F_2$.
In the HERA kinematical region the total $\gamma^* p$ cross-section is
related
to $F_2(x,Q^2)$ by
\begin{equation}
\sigma_{\rm tot}^{\gamma^* p} (s, Q^2) 
\simeq {4 \pi^2 \alpha \over Q^2} F_2 (x,Q^2) 
\end{equation}
where $s \simeq Q^2/x$ is the CMS energy squared 
for the $\gamma^* p$ collision.
For $Q^2 < 0.65$ GeV$^2$ and $s \geq 3$GeV$^2$ 
the $\sigma_{\rm tot}$ data \cite{h1,zeus99,zeus00}
seems to be well described by a 
combined Regge and Generalized Vector Meson Dominance (GVMD) 
motivated fit.
The ZEUS Collaboration used  the 4 parameter fit \cite{zeus00} 
\begin{equation}
\sigma_{\rm tot}^{\gamma^* p} (s, Q^2) =
\biggl( {M_0^2 \over M_0^2 + Q^2} \biggr)
\biggl( A_R s^{\alpha_R-1} + A_P s^{\alpha_P-1} \biggr)
\end{equation}
to describe the low $Q^2$ region, with
$A_R = 147.8 \pm 4.6 \mu$b, $\alpha_R=0.5$ (fixed),
$A_P =  62.0 \pm 2.3 \mu$b,
$\alpha_P = 1.102 \pm 0.007$ and $M_0^2 = 0.52 \pm 0.04$GeV$^2$. 
For $Q^2$ larger than 1 GeV$^2$ the HERA data on $F_2$ seems to
be well described by DGLAP evolution.
Parametrising $F_2 \sim A x^{- \lambda}$ at small $x$ 
the effective intercept $\lambda$ is observed to grow
from $0.11 \pm 0.02$ at $Q^2=0.3$GeV$^2$ 
to $0.18 \pm 0.03$ at $Q^2=3.5$GeV$^2$, $0.31 \pm 0.02$ at 35 GeV$^2$
and increases with increasing $Q^2$ \cite{zeus99,h1,desgrolard}.

What do we expect for $g_1$ ?

Let $\sigma_A$ and $\sigma_P$ denote the $\gamma p$ total cross-section 
for photons polarized antiparallel $\sigma_A$ and parallel $\sigma_P$ 
to the spin of the proton.
In HERA kinematics $g_1$ is related to $(\sigma_A - \sigma_P)$ by
\begin{equation}
\biggl(\sigma_A - \sigma_P \biggr) \simeq {4 \pi^2 \alpha \over p.q} g_1
\end{equation}
where $p$ and $q$ are the proton and photon four-momenta respectively.
The Regge prediction for the large $s_{\gamma p}$ 
behaviour of $(\sigma_A - \sigma_P)$ is \cite{heim,ek,clos1,sbpl,clos2}
\begin{equation}
\Biggl( \sigma_A - \sigma_P \Biggr) \sim
N_3 s^{\alpha_{a_1} - 1} + N_0  s^{\alpha_{f_1} - 1}
+ N_g {\ln({s / \mu^2}) \over s}
+ N_{PP} {1 \over \ln^2 ({s / \mu^2}) }
\end{equation}
at large $s$.
The $s^{\alpha_{a_1}-1}$ contribution is isotriplet; 
the $s^{\alpha_{f_1}-1}$, $(\ln s)/s$ and $1/\ln^2s$ 
contributions are isosinglet;
$\mu$ is a typical hadronic scale $\mu \sim 0.5 - 1$GeV.
The coefficients 
$N_3$, $N_0$, $N_g$ and $N_{PP}$ are to be determined from experiment.

If one makes the usual assumption that the $a_1$ and $f_1$
trajectories are straight lines running parallel to the 
$(\rho, \omega)$ trajectories,
then one finds 
$\alpha_{a_1} \simeq \alpha_{f_1} \simeq -0.4$.
This value is within the phenomenological 
range $-0.5 \leq \alpha_{a_1} \leq 0$ quoted by Ellis and Karliner \cite{ek}.
The $\ln s / s$ term is induced by any vector component to the short range 
exchange potential~\cite{clos1}.
It corresponds to a term in $g_1$ proportional to $\ln x$.
Extending the Donnachie-Landshoff-Nachtmann \cite{dln} 
model of soft pomeron exchange to $g_1$ Bass and Landshoff \cite{sbpl} 
found that the physics of nonperturbative two-gluon exchange, 
which generates the soft pomeron contribution to $F_2$, also 
generates a  $(2 \ln {1 \over x} -1)$ contribution in $g_1$. 
The $\ln x$ singularity is associated with the non-perturbative 
gluons emitted collinear with the proton.
The $1/\ln^2s$ term \cite{clos2} is associated with a two-pomeron cut.
Kuti \cite{kuti} has recently argued that the signature 
rule for Pomeron Regge cuts is expected to set $N_{PP}=0$.
In this paper we will allow this term to help parametrise any
less convergent behaviour of 
$(\sigma_A - \sigma_P)$ at large $s$ which might possibly 
show up in future data.

A high quality measurement of $(\sigma_A-\sigma_P)$ at $Q^2=0$ would help 
to constrain the high-energy part of the Drell-Hearn-Gerasimov sum-rule 
\cite{dhg,bassdhg} for spin dependent photoproduction.
This sum-rule 
relates the difference $(\sigma_A-\sigma_P)$ to the 
square of 
the anomalous magnetic moment of the target nucleon
and is often quoted in the target-rest frame:
\begin{equation}
({\rm DHG}) \equiv
- {4 \pi^2 \alpha \kappa^2 \over 2 m^2} =
\int_{\nu_{th}}^{\infty} {d \nu \over \nu} (\sigma_A - \sigma_P)(\nu).
\end{equation}
Here $\nu$ is the LAB energy of the exchanged photon, $m$ is 
the nucleon mass and $\kappa$ is the anomalous magnetic moment.
Phenomenological Regge based estimates \cite{bb,bianchi}
suggest that $25\pm 10 \mu$b (about 10\%) of the sum-rule may 
come from $\sqrt{s_{\gamma p}} > 2.5$GeV 
-- the highest energy of 
the present Bonn-Mainz Drell-Hearn-Gerasimov experiment.

It is an open question how far one can increase $Q^2$ away from 
the photoproduction limit
 and still trust Regge theory to provide an 
accurate description of $g_1$ in HERA kinematics.
It is well known \cite{reyal}
that small $x$ behaviour of the form 
$g_1 \sim x^{-\alpha}$ where $\alpha < 0$ 
is unstable to DGLAP evolution \cite{grsv,gehr,abfr} 
and to resummation of $(\alpha_s \ln^2 {1 \over x})^k$ 
\cite{bart,blum,kiyo,ziaja} terms at small $x$ in perturbative QCD.
Theoretical studies suggest that the precise shape of $g_1$ at 
small $x$ in deep inelastic scattering is particularly sensitive 
\cite{ziaja}
to the details of the QCD evolution, and might even rise as fast
 as $| g_1 | \sim {1 \over x}$ \cite{bart}.

Regge theory provides a good fit \cite{donn} to the NMC 
fixed target experiment ``small $x$'' 
data \cite{nmc}
($0.008 < x < 0.07$) on $F_2^{(p \pm n)}(x,Q^2)$ at deep inelastic 
$Q^2$ between 1 and 10 GeV$^2$.
It does not appear to describe $g_1$ data in the same kinematical 
region.
Taking $\alpha_{a_1} \simeq -0.4$ in Eq.(4) yields a Regge prediction 
$g_1 \sim x^{0.4}$ at small $x$.
In contrast, polarized deep inelastic data from CERN \cite{smcq2}
and SLAC \cite{e143a,e154} 
consistently indicate a strong isotriplet term in $g_1$ which 
rises at ``small $x$'' between 0.01 and 0.1 at $Q^2 \simeq 5$GeV$^2$.
One finds a good fit \cite{soff,bb}
\begin{equation}
g_1^{(p-n)} \sim (0.14) \ x^{- 0.5}
\end{equation}
to the SLAC $g_1$ data which has the smallest experimental errors.
This fit corresponds to an effective intercept 
$\alpha_{a_1}(Q^2) \simeq +0.5$ for this kinematical 
region 
--- between 0.5 and 1.0 greater than the soft Regge 
prediction if we take the phenomenological range
$-0.5 \leq \alpha_{a_1} \leq 0$ \cite{ek}.
It is interesting to note that a large ``small $x$'' 
contribution to $g_1^{(p-n)}$ in this kinematics is 
almost necessary \cite{bass99} to accommodate 
the large area under the Bjorken sum-rule for $g_1^{(p-n)}$ \cite{bj}
--- a non-perturbative constraint. 
It will be interesting to see how well the fit (6) describes
future $g_1$ data at smaller Bjorken $x$.
For $g_1^{(p+n)}$ the situation is less clear.
The SLAC data indicates that $g_1^{(p+n)}$ is small and
consistent with zero in the $x$ range $0.01 < x < 0.05$.
Theoretically, one expects here a sum of 
different exchange contributions with possibly different signs.

To summarise, the change in the effective intercepts for $g_1^{(p \pm n)}$ 
between photoproduction and deep inelastic scattering, 
say at $Q^2 \sim 5$GeV$^2$, could 
be as large as one --- a factor of 5 bigger than the effect observed in $F_2$.

\section{Experimental aspects}

The lowest $Q^2$ values at large  $1/x$
can be reached in collider
type of lepton-hadron experiments. HERA is presently the only 
high energy $ep$ collider, with a 27.5 GeV electron and 820 (920) GeV
proton beam, leading to interactions with $\sqrt{s}= 300$ (314) GeV.
Presently the collider experiments at HERA record unpolarized $ep$ 
collisions. In fall 2000 spin rotators will be installed 
in the electron ring converting the transverse polarization  
of the electron beam, which builds up due to the 
Sokolov-Ternov~\cite{sokolov} effect,
 into  a physics-wise more interesting longitudinal polarization. 
Studies are 
being made to provide in future also a polarized
 proton beam at HERA ~\cite{barber}
which would enable polarized $ep$ and thus also polarized $\gamma p$ 
collisions.
In 2000 HERA will also undergo a luminosity upgrade~\cite{willeke}, 
which for the 
experiments has the consequence that magnets are inserted close to 
the interaction point, and the beam-line and beam optics will change.
In this mode HERA is expect to deliver a luminosity in the range of
150-200 pb$^{-1}$/year per experiment.

At BNL  a new project, called eRHIC, is under study\cite{erhic}.
It is proposed  to add 
a polarized electron ring/accelerator to the already existing 
and recently commissioned $pp$/$AA$
machine RHIC. Polarized proton beams are already planned for RHIC.
Hence
polarized $ep$ collisions at a CMS energy of about 100 GeV will be 
possible at eRHIC (10 GeV $e$ on 250 GeV $p$).

If HERA is fully polarized, event samples of the order of 100 $-$ 500 
pb$^{-1}$ are expected to be collected, with expected beam polarizations 
$P_e =  P_p =  0.7 $. For eRHIC we will assume the same beam polarizations
but a higher total integrated luminosity, namely of order of 
 1 fb$^{-1}$/year.

As discussed above, in spin physics 
non-singlet and singlet contribution  decomposition is 
important, and for that reason it 
is crucial  to have also access to the spin structure functions of the neutron.
The original idea at HERA was to use additionally He${^3}$ beams, 
which strongly resemble the protons
from the spin acceleration point of view.
Recently, the idea to use deuterons has re-emerged. High energy
polarized deuteron beams have in fact 
several advantages compared to protons. Due to the much smaller
gyromagnetic anomaly $G = (g-2)/2$ (1.79 for protons, $-0.14$ for
deutrons)
it will be easier to accelerate a polarized deutron beam over the 
depolarizing resonances, and the beams are less susceptible to 
spin distortions\cite{deuterons}. The known disadvantage is that with 
current magnet technology it 
is not possible to use spin rotators to rotate the transverse 
deuteron spin into a 
longitudinal one. Novel ideas for rotating the spin based on 
magnetic $rf$ dipole fields could however change this situation
significantly.
Here we will assume that  high energy 
polarized deuteron beams can be made and stored (in HERA up to 460 GeV/nucleon
and eRHIC up to
125 GeV/nucleon), with the same luminosity and same polarization
as for protons (which does not seem unfeasible~\cite{zelenski}).
Note that if tagging of the spectator particle,
 i.e. the particle which did not undergo an interaction,
with high efficiency and kinematical
coverage could be achieved, by means of  proton spectrometers or 
neutron calorimeters down the beamline (also termed 
``the forward direction''), one could measure directly both 
 $ep$ and $en$ scattering contributions separately using event samples
with either a proton or neutron spectator tagged. Hence one can measure 
separately but simultaneously $g_1^p$ and $g_1^n$, as already proposed
in ~\cite{duren}. Some coverage for the detection 
of forward protons and neutrons is already
available at HERA.
An excellent, quasi-complete 
coverage of the forward direction is foreseen in the first 
ideas for a detector at eRHIC~\cite{krasny}.

\subsection{Photoproduction}

Photoproduction, i.e. $Q^2= 0$,  cross-sections can be measured in $ep$ 
collisions at a collider at high 
$\sqrt{s_{\gamma p}}$ energies. In fact the dominant processes in 
$ep$ collisions are $\gamma p  $ interactions where the photon is on
mass shell. The electron is scattered under approximately zero degrees
with respect to the electron beam direction, 
which means that  it remains
in the beampipe. The energy of the scattered electron 
$E_e'$ is however reduced to $E_e' = E_e -E_{\gamma}$, with $E_e$
the incident electron  energy and $E_{\gamma}$  the emitted photon energy.
The HERA machine magnets in the beamline, 
which steer the beam into a closed orbit, 
act as a spectrometer on these off-momentum electrons, and they will be 
kicked out of the beam orbit.
The experiments H1 and ZEUS  
have installed calorimeters to detect these kicked out electrons along
the beamline.
In case of H1 calorimeters (stations) 
are installed at three locations: at 8 m , 30 m  and 44 m
distance from the interaction point~\cite{H1}.
 The stations accept (tag) electrons from 
different momentum ranges, which correspond to $\sqrt{s_{\gamma p}}$ ranges 
of 280-290 GeV, 150-250 GeV and 60-115 GeV respectively.
At the central energy value of each region the acceptance of these devices 
amounts to 20\%, 80\% and 70\% respectively.

Eq.(\ref{sigtot}) below 
shows that the $\gamma p$ cross section is large, of order
of hundreds of microbarns. The photon energy spectrum 
emitted from an electron beam follows the
 Weizs\"acker-Williams approximation~\cite{www}. 
An integrated $ep$ luminosity 
at HERA of 1 pb$^{-1}$ can yield about $ 1500$K $ \gamma p$ events
in each of the 30 m and 44 m stations, and about 10 times less in the 8 m
station.

In the small asymmetry approximation the 
error on the asymmetries, $\delta A$, can be calculated as
$1/(P_eP_p\sqrt{N})$.  Due to data-taking bandwidths and 
trigger challenges presently not all tagged $\gamma p$ events 
are recorded. Assuming a data taking rate of 2 Hz for these events,
 also in 
future, leads to  about 40 M events/year
giving  a reachable precision on the $ep$ asymmetry
  $\delta A = 1/(P_eP_p\sqrt{N})= 0.0003$.
It is however not excluded that novel techniques in 
triggering, data-taking and on-line analysis
will become available, which would allow to
collect or use the information of all produced events, amounting to
approximately
15,000M events in total for  the 
three stations in a period of 3 to 5 years. This would lead to  
 maximal reachable 
precisions of $\delta A = 3.10^{-5} $
for a measurement at the 30m and the 44m station, and $\delta A = 10^{-4} $
for a measurement at the 8m station.

 Since the $\gamma p$ asymmetries are measured via $ep$ 
collisions the polarization of the photon beam 
will be reduced by a so called depolarization
factor $D = y(2-y)/(y^2+2(1-y))$
with  $y = s_{\gamma p}/s_{ep}$. For the three stations the measurements
are at $y= 0.09, 0.44$ and 0.90, leading to values of $ D = 0.094, 0.52$
and 0.98. 
Hence the maximal reachable precision for measuring the  
$\gamma p$ asymmetries $A_1$
 using $ep$ at HERA is $10^{-4}$ (280-290 GeV), $6\cdot 10^{-5}$
(150-250 GeV) and 
$3\cdot 10^{-4}$ (60-115 GeV).

If the same photoproduction tagging techniques as used at
HERA can be used at  eRHIC, and if the data taking 
and analysis bandwidth is not 
preventative, an experiment at this machine could measure the 
$\gamma p$ cross section asymmetry with a precision of 
about a factor three better, in the region $20 <
\sqrt{s_{\gamma p}} < 90$ GeV,
complementary to HERA.

Using a deuteron beam one could measure the $A^{p+n}_1$ asymmetry 
with the same precision, but for reduced $\sqrt{s_{\gamma p}}$ regions, namely
$40-200$ GeV at HERA and $14-70$ GeV
at eRHIC. Using spectator tagging the
individual 
$A^p_1$ and $A^n_1$ asymmetries can be 
measured with roughly a factor of two worse precision, compared to
$A_1^{p+n}$.

After the luminosity upgrade of HERA
the present 
electron taggers for photoproduction events will undergo changes, 
be partially located at different positions, and have a different 
acceptance. The regions which are expected to be covered after the 
upgrade for $Q^2=0$ events are 
%$0.05 < y<0.25$ and $0.5<y<0.78$
$70 < \sqrt{s_{\gamma p}}< 150 $  GeV and 
$210 < \sqrt{s_{\gamma p}}< 265 $ GeV, but the precision to measure
the 
asymmetries should be as above.

\subsection{Low $Q^2$ Region.}

The low $Q^2$, or transition, region from 
photoproduction to deep inelastic scattering is 
investigated presently at the HERA collider for unpolarized $ep$ collisions 
by the H1 and ZEUS experiments. Similarly to the photoproduction
events discussed above, 
the main characteristic of low $Q^2$ scattering in the HERA LAB frame is
the 
small but this time non-zero scattering angle of the electron with 
respect to the beam direction. 
These electrons will leave the beampipe a few meters after 
the interaction point.
Hence special care has to be taken to detect these electrons with 
detectors which are closely integrated with the beam-pipe structure of the 
accelerator. 

\begin{figure}[t]
\begin{center}
\begin{picture}(500,200)(0,0)
\epsfig{file=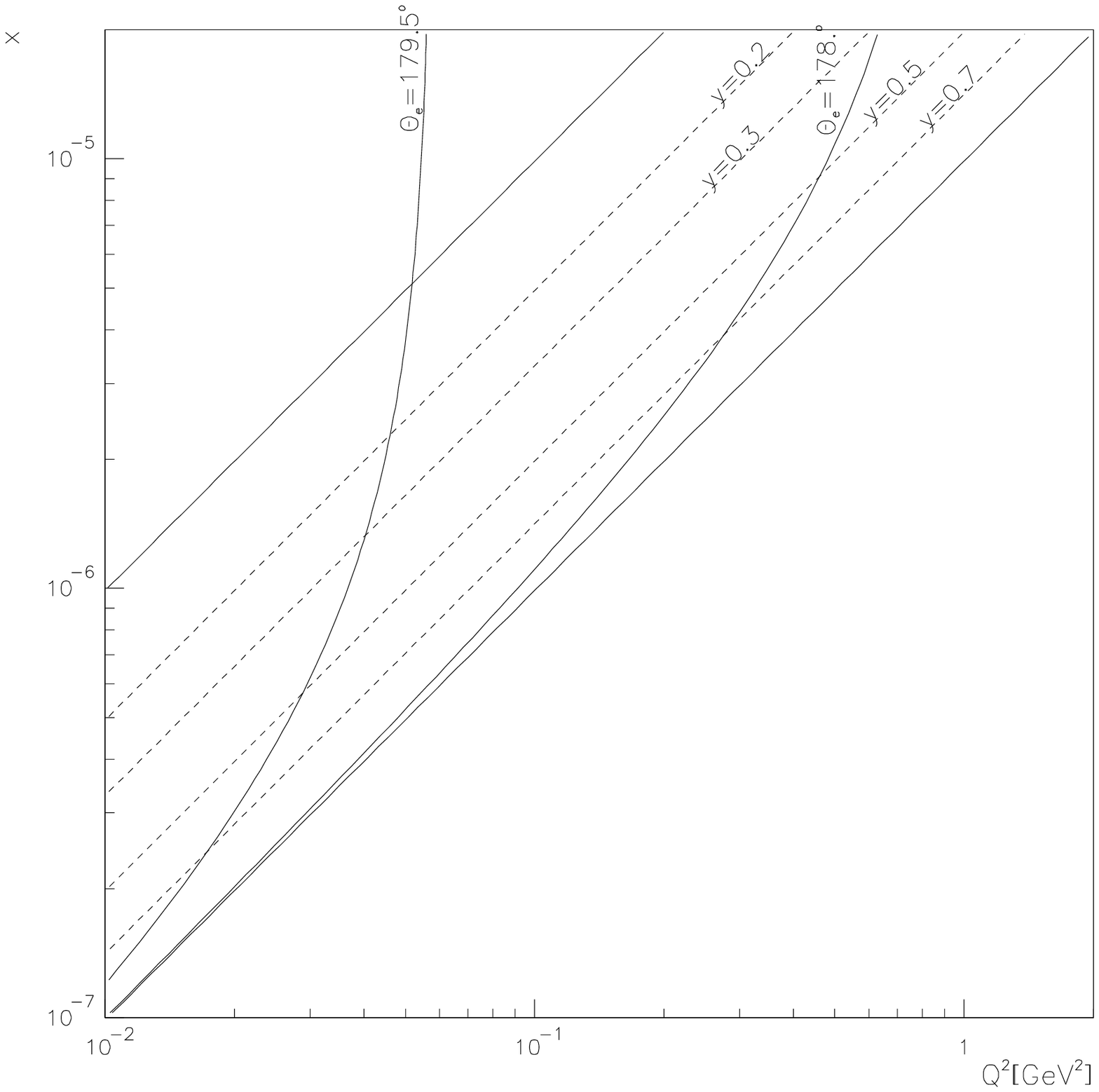,width=10cm,bbllx=-150pt,bblly=150pt,
bburx=410pt,bbury=700pt}
\put(-140,200){\bf a)}
\end{picture}

\vspace{3cm}
\begin{picture}(500,200)(0,0)
\epsfig{file=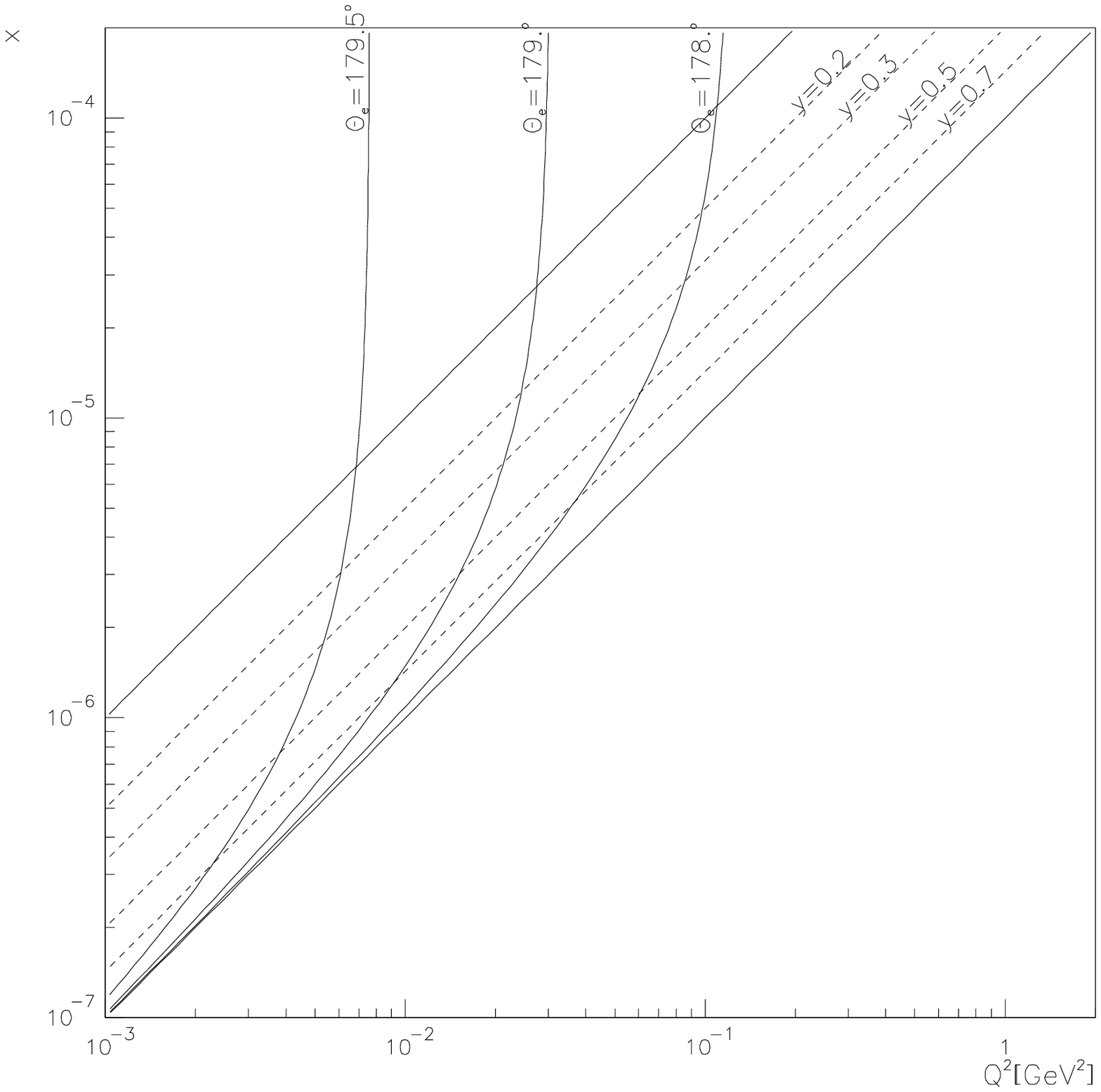,width=10cm,bbllx=-150pt,bblly=150pt,
bburx=410pt,bbury=700pt}
\put(-140,200){\bf b)}
\end{picture}
\end{center}
\caption{The kinematical region accessable by {\bf a)} HERA and
{\bf b)} eRHIC.
The angles are defined with respect to the proton beam direction.}
\label{fig:kinem}
\end{figure}

Fig.\ref{fig:kinem}a) shows the kinematics of the scattered electron 
for HERA. In order to reach small $Q^2$ values, below 0.1 GeV$^2$,
electrons 
need to be detected with scattering angles of $1^0$ or less.

The standard central detectors of H1 and ZEUS typically detect electrons down 
to an angle of about 2 degrees, 
resulting in  a region of good acceptance for $Q^2$ values
above a few GeV$^2$.
In order to reach smaller values both experiments have added to the central 
detector a special ``beam-pipe calorimeter'' and ``tracker'' 
(BPC and BPT; ZEUS) 
and ``Very Low $Q^2$'' detector (VLQ; H1). 
These detectors have however a limited azimuthal coverage.
ZEUS has already used the
BPC and BPT detectors to present measurements of $F_2$ based on 1997
data.
The measurements  cover the region of $Q^2$ down to 
0.045 GeV$^2$~\cite{lowq2zeus}.
The VLQ detector of H1~\cite{vlq} consists of a silicon tracker and 
 Tungsten-scintillator calorimeter and was commissioned in 1998. 
The VLQ is expected to 
provide measurements down to 0.03-0.05
 GeV$^2$.
In this paper we 
will assume one can detect electrons down to 10 mrad, which is slightly
better than presently available with the VLQ and BPT.
With present standard methods for event kinematics reconstruction the region 
in Bjorken $y$ of $0.1 < y < 0.7$ can be safely measured. 
 Further we will restrict the measured $Q^2$  region from 
0.03 GeV$^2$ to 1.5 GeV$^2$ in $Q^2$.
Resolution and event statistics considerations for polarized
measurements lead to the possible binning 
in $Q^2$ and $s_{\gamma p}$ given in Table 1.
Using the parametrization of ~\cite{zeus00} in this region, 
the number of events per bin is
of the order of 
$1-2\cdot 10^6$  for an integrated 
luminosity of 100 pb$^{-1}$.
One interesting aspect to be studied with the data is the energy dependence
of the unpolarized cross section. With these data samples and assuming 
that the 
systematic errors on the measurements will be mostly correlated,
one will measure the exponents $\alpha_R, \alpha_P$ 
in Eq.(2) with an unprecedented precision of
better than 10$^{-3}$.

The luminosity upgrade at HERA will
effectively
reduce the acceptance for scattered electrons for low $Q^2$ DIS events,
due to the focussing magnets in the detectors,
 close to the interaction
point,
  and
 measurements in the kinematical range presently covered by the  VLQ
are not foreseen. 
Hence, for the low $Q^2$ measurement 
a new way of detecting and measuring DIS events within the 
upgraded environment has to be 
 found, in order that  this measurement can 
benefit from the luminosity upgrade, which would give approximately
 150 
pb$^{-1}$/year.
Without these focussing 
magnets (but still some less aggressive improvements compared
to the present luminosity upgrade~\cite{brinkman}) HERA could 
still accumulate 
between 50 and 100 pb$^{-1}$/year.
We assume here  that this measurement will or can be made
at a later stage in a
lower luminosity configuration, and assume a total luminosity  
of 100pb$^{-1}$,  
corresponding to 1-2 years of data taking. 
This leads to a reachable precision 
$\delta A = 1/(P_eP_p\sqrt{N})= 0.002-0.001$ per bin.
When less bins are chosen and/or a solution is found to make this 
measurement within
the upgraded luminosity program, the ultimate 
sensitivity can increase with a factor of 3 to 4.
Note that the true $\gamma^* p$ asymmetries are reduced due to
the depolarization factor $D$, which lowers the
effective  sensitivities by 
 values which  range from  0.17 to  0.72. Hence we take 
$\delta A_1 = 0.002$ as a typical reachable value for the 
precision of the $\gamma^*p$ asymmetry measurement.

When running HERA at lower energies (e.g. $E_p$ = 200 GeV and $E_e$ = 15 GeV),
one could also reach $s_{\gamma p}$ values which are a factor of 3 lower 
than the ones in Table 1.
Note however that the sensitivity will reduce by about a factor 5 to 10 or 
so, due to the reduced luminosity of HERA at these energies, and the fact 
that it is unlikely that data at such energies would be taken for a full year.

For eRHIC access to the low $Q^2$ region becomes easier, as
shown in Fig.\ref{fig:kinem}b). Due to the 
different beam energies electrons from DIS events with a $Q^2$
of 0.1 GeV$^2$ scatter at an angle of 2 degrees or more, which allows
them to be accepted within the central detector. Special beam-pipe 
calorimeters and trackers are only required for the region $0.01< Q^2< 0.1$
GeV$^2$. If also at eRHIC scattered
electrons can be tagged down to 10 mrad, then 
$Q^2$ values down to 0.003 GeV$^2$ can be reached.
The higher luminosity of RHIC and larger azimuthal acceptance will 
allow one to measure the asymmetries 10-15 times more precisely, but in 
a kinematic range where, for the same $Q^2$, 
 $x$ is 10 times larger than for HERA.

As for the photoproduction case,
using a deuteron beam one could measure the $A^{p+n}_1$ asymmetry 
with the same precision, but for reduced $s_{\gamma p}$ regions, 
i.e. for $x$ values a factor two larger than those accessible with 
proton beams for the same $Q^2$.
In case spectator tagging is available  
$A_1^p$ and $A_1^n$ asymmetries could be measured simultaneously
with a factor of about two worse precision.

\section{Estimating asymmetries}

We now estimate the spin asymmetry 
$A_1 = (\sigma_A - \sigma_P) / (\sigma_A + \sigma_P)$ at low $Q^2$ 
and discuss the measurement potential of Polarized HERA and eRHIC.

\subsection{Photoproduction}

In Fig.2 we show the estimate of the real photon asymmetry $A_1$ 
from \cite{sma,bb}.
This estimate was obtained as follows.
We took the SLAC E-143 \cite{slacq2} and SMC \cite{smcq2} 
measurements of 
$A_1 = (\sigma_A - \sigma_P) / (\sigma_A + \sigma_P)$ 
at low $Q^2$ 
(between 0.25 and 0.7 GeV$^2$) 
in the ``Regge region'' ($\sqrt{s_{\gamma p}} \geq 2.5$GeV). 
This low $Q^2$ data exhibits 
no clear $Q^2$ dependence in either experiment.
Motivated by this observation and 
the ZEUS fit (2) to $F_2(x,Q^2)$ at low $Q^2$
we assume that $A_1$ is $Q^2$ independent for $Q^2 < 0.7$GeV$^2$.
That is, we conjecture
\begin{equation}
(\sigma_A - \sigma_P)^{\gamma^* p}(s,Q^2) = 
\biggl( {M_0^2 \over M_0^2 + Q^2} \biggr) 
\ 
(\sigma_A - \sigma_P)^{\gamma p}(s,0)
\end{equation}
at large $s_{\gamma p}$ and small $Q^2$ 
with the same value of $M_0^2$ in both Eqs.(2) and (7). 
The SLAC and SMC low $Q^2$ data was combined to obtain one proton and 
one deuteron point corresponding to each experiment. 
The combined SLAC data at low $x$ and low $Q^2$ exhibits a clear 
positive proton asymmetry
$A_1^p = +0.077 \pm 0.016$ at 
$\langle \sqrt{s_{\gamma p}} \rangle = 3.5$GeV and 
$\langle Q^2 \rangle = 0.45$GeV$^2$
while the deuteron asymmetry $A_1^d = +0.008 \pm 0.022$ 
is consistent with zero. 
The small isoscalar deuteron asymmetry $A_1^d$ 
indicates that the isoscalar contribution 
to $A_1^p$ in the E-143 data is unlikely to be more than 30\%.
At larger $\langle \sqrt{s_{\gamma p}} \rangle = 16.7$GeV
the
SMC proton and deuteron low $Q^2$ asymmetries are both 
consistent with zero:
$A_1^p = +0.011 \pm 0.013$ and $A_1^d = +0.002 \pm 0.014$ at a
mean 
$\langle Q^2 \rangle = 0.45$GeV$^2$.
For the total photoproduction cross-section we took
\begin{equation}
(\sigma_A + \sigma_P) 
= 67.7 \ s_{\gamma p}^{+0.0808} + 129 \ s_{\gamma p}^{-0.4545}
\label{sigtot}
\end{equation}
(in units of $\mu$b),
which is known to provide a good Regge fit for $\sqrt{s_{\gamma p}}$ 
between 2.5GeV and 250GeV~\cite{pvl1}.
(Here, 
the $s_{\gamma p}^{+0.0808}$ contribution is associated with pomeron
exchange
and the $s_{\gamma p}^{-0.4545}$ contribution is associated with the 
isoscalar $\omega$ and isovector $\rho$ trajectories.)
Assuming a $Q^2$--independent $A_1$ we combined Eqs.(4) and (8)
to make various Regge fits through the SLAC proton point,
which exhibits the only significant signal among the SLAC and SMC 
proton and deuteron
low $x$ low $Q^2$ asymmetries.
For definiteness we take $\mu^2=0.5$GeV$^2$ in (4).

In Fig.~2 we show the asymmetry $A_1^p$ as a function of 
$\sqrt{s_{\gamma p}}$ between 2.5 and 250 GeV
for the 
four different would-be Regge behaviours for
$(\sigma_A - \sigma_P)$:
that the high energy behaviour of $(\sigma_A-\sigma_P)$ 
is given
\begin{enumerate}
\item
entirely by the $(a_1, f_1)$ terms in Equ.(4) with 
Regge intercept $-{1 \over 2}$ (conventional) 
\item
entirely by the $(a_1, f_1)$ terms in Equ.(4) with 
Regge intercept $+{1 \over 2}$
  (motivated by the observed small $x$ behaviour of $g_1^{(p-n)}$ in Eq.(6)),
\item
by taking 2/3 isovector (conventional) $a_1$ and 
1/3 two non-perturbative 
gluon exchange contributions at $\sqrt{s_{\gamma p}} = 3.5$GeV,
\item
by taking 2/3 isovector (conventional) $a_1$ and 1/3 
pomeron-pomeron cut
contributions at $\sqrt{s_{\gamma p}} = 3.5$GeV.
\end{enumerate}

Polarized HERA could measure the real-photon spin asymmetry $A_1$ 
for $\sqrt{s_{\gamma p}}$ between 60 and 280 GeV 
to precision 
$\delta A_1 \simeq 0.0003 - 0.0001$;
eRHIC to precision $\delta A_1 \simeq 0.0001-0.00003$ 
for $\sqrt{s_{\gamma p}}$ between 20 and 90 GeV --- see Section 3
and \cite{sma}.
This polarized photoproduction measurement will help to constrain 
our understanding of spin-dependent Regge theory and to put an upper 
bound on the high-energy part of the Drell-Hearn-Gerasimov sum-rule.
Given the projected asymmetries, 
Polarized HERA with $\delta A_1 = 0.0003$ would be sensitive 
to $(\sigma_A-\sigma_P)$ falling no faster than about 
$s_{\gamma p}^{-1}$ ($\alpha_{a_1} = 0$) at $\sqrt{s_{\gamma p}}=60$GeV.
For
eRHIC with $\delta A_1 = 0.0001$ 
one is sensitive to
$(\sigma_A-\sigma_P)$ 
falling no faster than about $s_{\gamma p}^{-1.9}$ 
($\alpha_{a_1} = -0.9$)
at the lower energy $\sqrt{s_{\gamma p}}=20$GeV, 
which is well within all theoretical expectations 
for the large $s_{\gamma p}$ behaviour of the asymmetry.
At the upper energy
$\sqrt{s_{\gamma p}}=90$GeV one expects to see a signal
for
$(\sigma_A-\sigma_P)$ falling no faster than about $s_{\gamma p}^{-1}$ 
($\alpha_{a_1} = 0$).

\subsection{Low $Q^2$ Region}

Polarized eRHIC and HERA could measure the $\gamma^* p$ spin asymmetry 
in the transition region 
($0.05 < Q^2 < 1$GeV$^2$) 
to precision $\delta A_1 \simeq 0.00015$ and $0.002$ respectively  --- 
see Section 3.
At eRHIC the values of $x$ are typically 10 times larger than for HERA 
for the same value of $Q^2$.

In Table 1 we estimate the size that $g_1$ has to be in order to see a 
signal in each of practical $(\sqrt{s_{\gamma p}},Q^2)$ bins
for the Polarized HERA and eRHIC Colliders.
In a first approximation we take $F_L = 0$ and use the ZEUS fit, 
Eq.(2), 
to $F_2$ at low $Q^2$ to calculate $g_1 = \delta A_1 \ F_2 / 2x$ \
\footnote{
Errors on $F_2$ for HERA and eRHIC will be typically of the order of a 
few \%, making $\delta A_1$ by far the major source of experimental
 error.}.
For $\delta A_1$ we take the values as determined in Section 3.
These $g_1$ values are about a factor 10 (HERA) or even 100 (eRHIC) 
smaller
  the estimations  for $g_1$ 
     at 10GeV$^2$ 
     in the same range of $x$, 
     which follow from extrapolating a QCD fit \cite{epjc}
 to present $g_1$ data from DESY, SLAC and SMC.
Hence if the asymmetries follow the perturbative predictions
down to $Q^2$ =1 GeV$^2$, both colliders will observe measurable
asymmetries already after a small fraction of the data samples 
are collected.

\begin{table}[h]
\footnotesize
\begin{center}
\begin{tabular}{||c|r|r|r|r|r||} \hline \hline
\multicolumn{6}{||c||}{} \\
\multicolumn{6}{||c||}{$Q^2$} \\
\multicolumn{6}{||c||}{} \\
\hline
             & $0.05$   & $0.1$  & $0.2$ & $0.5$ & $1.0$
\\
\hline
\multicolumn{6}{||c||}{} \\
\multicolumn{6}{||c||}{Polarized HERA with $\delta A_1 = 0.002$} \\
\multicolumn{6}{||c||}{} \\
\hline
$\sqrt{s_{\gamma p}} \ (y)$    &          &        &       &       & 
\\
\hline
120  (0.16)  & (3.5x10$^{-6}$, 19)   &  (6.9x10$^{-6}$, 18)  & 
               (1.4x10$^{-5}$, 15)   &  (3.5x10$^{-5}$, 11)  & 
               (6.9x10$^{-5}$, \  7)   \\
\hline
150  (0.25)  & (2.2x10$^{-6}$, 31)   &  (4.4x10$^{-6}$, 28)  & 
               (8.9x10$^{-6}$, 24)   &  (2.2x10$^{-5}$, 17)  & 
               (4.4x10$^{-5}$, 12)   \\
\hline
190  (0.4)   & (1.4x10$^{-6}$, 51)   &  (2.8x10$^{-6}$, 47)  & 
               (5.5x10$^{-6}$, 41)   &  (1.4x10$^{-5}$, 29)  & 
               (2.8x10$^{-5}$, 20)   \\
\hline
230  (0.6)   & (9.5x10$^{-7}$, 79)   &  (1.9x10$^{-6}$, 73)  & 
               (3.8x10$^{-6}$, 63)   &  (9.5x10$^{-6}$, 45)  & 
               (1.9x10$^{-5}$, 30)   \\
\hline
\multicolumn{6}{||c||}{} \\
\multicolumn{6}{||c||}{Polarized eRHIC with $\delta A_1 = 0.00015$} \\
\multicolumn{6}{||c||}{} \\
\hline
38 (0.14)    & (3.5x10$^{-5}$, 0.12)   &  (6.9x10$^{-5}$, 0.11)  & 
               (1.4x10$^{-4}$, 0.09)   &  (3.5x10$^{-4}$, 0.07)  & 
               (6.9x10$^{-4}$, 0.04)   \\
\hline
47.5 (0.23)        & (2.2x10$^{-5}$, 0.19)   &  (4.4x10$^{-5}$, 0.18)  & 
               (8.9x10$^{-5}$, 0.15)   &  (2.2x10$^{-4}$, 0.11)  & 
               (4.4x10$^{-4}$, 0.07)   \\
\hline
60  (0.36)         & (1.4x10$^{-5}$, 0.32)   &  (2.8x10$^{-5}$, 0.29)  & 
               (5.5x10$^{-5}$, 0.26)   &  (1.4x10$^{-4}$, 0.18)  & 
               (2.8x10$^{-4}$, 0.12)   \\
\hline
72.5 (0.53)        & (9.5x10$^{-6}$, 0.48)   &  (1.9x10$^{-5}$, 0.44)  & 
               (3.8x10$^{-5}$, 0.38)   &  (9.5x10$^{-5}$, 0.27)  & 
               (1.9x10$^{-4}$, 0.18)   \\
\hline \hline
\end{tabular}
\caption{\em 
The values $(x, g_1 = \delta A_1 F_2 / 2x)$ 
for each of the practical $(\sqrt{s_{\gamma p}},Q^2)$ bins for Polarized
HERA and eRHIC.
Note that $x$ is a factor of 10 higher for each $Q^2$ bin compared to the 
HERA binning. 
\label{int}}
\end{center}
\end{table}

Depending on the size of $g_1$ at these low values of $x$, 
which remains to be measured,
it is very likely to be possible to observe how the effective intercepts 
for 
$g_1^{(p-n)}$ and $g_1^{(p+n)}$ evolve with $Q^2$ in the lower $Q^2$ 
region.
If the spin asymmetry $A_1$ is indeed $Q^2$ independent up to 
$Q^2 \simeq 0.5$GeV$^2$, 
then taking the asymmetry estimates in Fig. 2 one would expect to see 
a signal with eRHIC at 
$\sqrt{s_{\gamma p}} \simeq 38$ GeV 
if $g_1$ is less convergent than about 
$g_1 \sim x^{0.3}$ as $x \rightarrow 0$ 
with fixed $Q^2 < 0.5$GeV$^2$. 
For Polarized HERA with $\delta A_1 = 0.002$
there will, most likely,
be no definite signal 
if $g_1$ follows the Regge behaviour $g_1 \rightarrow 0$ 
when $x \rightarrow 0$ at fixed low $Q^2$.
If $|g_1|$ at low fixed-$Q^2$ rises at small $x$, 
possibly due to significant 
gluonic $\ln x$ or $1/x \ln^2 x$ contributions, 
one could expect a measurable low $Q^2$ asymmetry in both colliders.
Experimentally, the strategy should be to measure 
$A$ at low $x$ with decreasing $Q^2$ until the 
asymmetry becomes too small to be significant.
The further one can probe into the transition region, the greater 
the constraints one can provide on models of the Regge to hard $Q^2$ 
transition at small $x$.

\section{Conclusions}

Exploration of the transition region between polarized photoproduction and 
deep inelastic scattering, $0 < Q^2 < 1$GeV$^2$,
looks feasible with Polarized HERA and eRHIC. 
Polarized photoproduction measurements at these colliders would constrain 
our knowledge of spin dependent Regge theory and put an upper bound on 
the high-energy Regge contribution to the Drell-Hearn-Gerasimov sum-rule.
Much larger changes in the effective intercept for the spin dependent
structure function $g_1$ at small $x$ than for spin independent $F_2$
are expected.
A dedicated measurement at Polarized eRHIC or HERA 
would impose new constraints on QCD based models of the 
transition from Regge theory to perturbative QCD with increasing $Q^2$ at 
fixed low $x$.

HERA and eRHIC will cover complementary regions in kinematics for these
measurements, and will thus both provide important information. 
The potential high luminosity at eRHIC is certainly an
advantage, and allows  to reach high sensitivities in the transition
region. Futhermore, it may be experimentally easier to reach 
lower $Q^2$ with eRHIC. HERA will need again dedicated detectors
to access the low $Q^2$ region after the luminosity upgrade. 
With appropriate small angle tagging detectors in both the 
electron and proton direction, very precise data on asymmetries
can be collected, which will be important in the progress 
of this data driven field to the study of the strong interaction.
Hence, having such detector coverage included in design of
a new detector for measuring $ep$ and $eA$ collisions at eRHIC is
strongly encouraged.
The use of deuteron beams with spectator tagging can help to
disentangle different exchange contributions in the Regge regime.

\vspace{1.0cm}

{\bf Acknowledgements} \\

{\noindent 
It is a pleasure to thank B. Badelek, D. Barber,
A. Deshpande, G. Hofstatter, S. Levonian and G. R\"adel for helpful
discussions.
SDB acknowledges the hospitality of the CERN Theoretical Physics 
Division where this work was begun. 
This work was supported in part by BMBF and DFG.
}

\vspace{1.0cm}

%\newpage

\newpage

\begin{figure}
\centerline{}
\begin{center}
%\begin{flushright}
\input{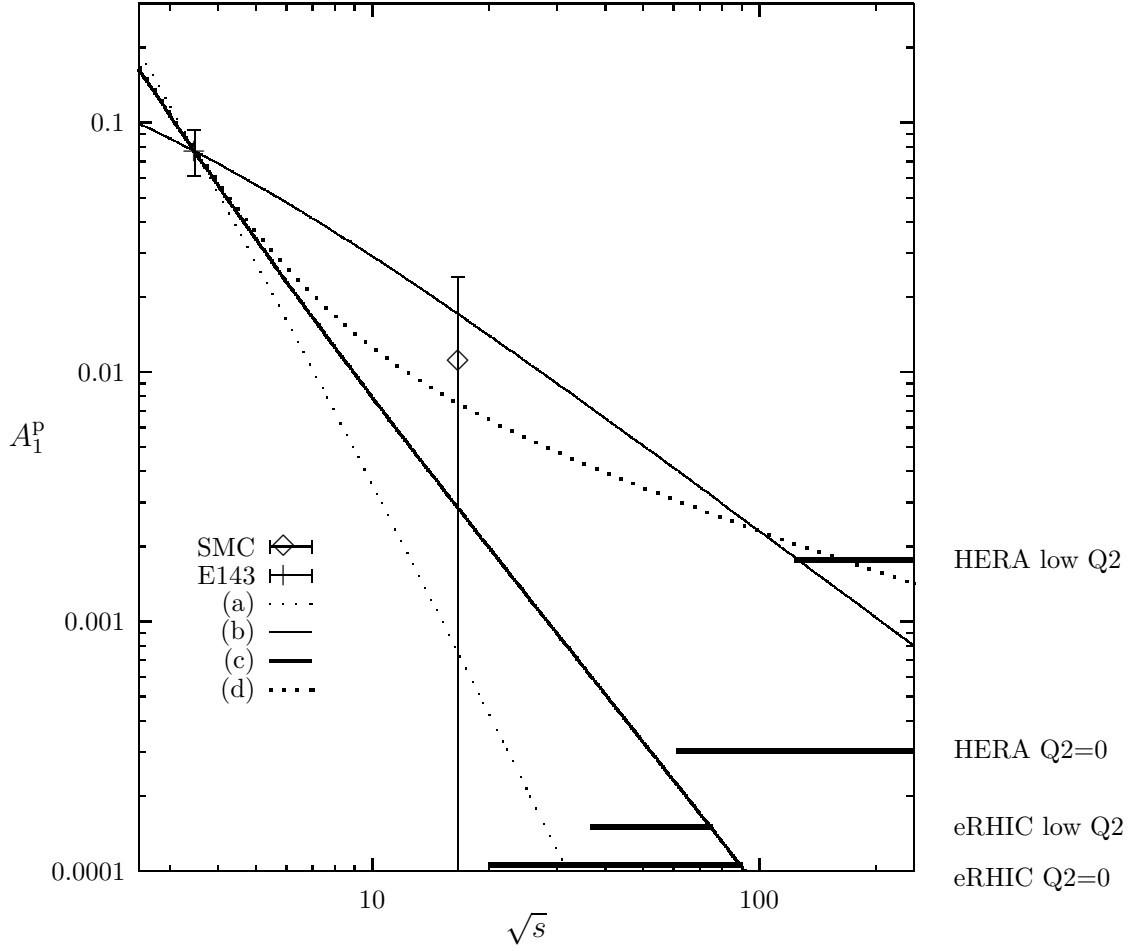}
%\end{center}
\caption{
{\bf The asymmetry $A_1^p$}
as a function of $\sqrt{s_{\gamma p}}$ for different Regge behaviours 
for
$(\sigma_A - \sigma_P)$:
 given
entirely by (a) the $(a_1, f_1)$ terms in Eq.(4) with Regge 
    intercept either 
    $-{1 \over 2}$ (conventional) or
    (b)  $+{1 \over 2}$
(c) by  2/3 isovector (conventional) $a_1$ and 
1/3 two non-perturbative gluon exchange contributions at 
$\sqrt{s} = 3.5$GeV; 
(d)
by 2/3 isovector (conventional) $a_1$ and 1/3 
pomeron-pomeron cut
contributions at $\sqrt{s} = 3.5$GeV.
The solid horizontal lines indicate the kinematic range and 
precision of the Polarized eRHIC and HERA colliders at low $Q^2$.
}
%\end{flushright}
\end{center}
\end{figure}

\end{document}